\documentclass[a4paper,11pt]{article}
\usepackage{pos}
\usepackage{setspace}

\definecolor{ao(english)}{rgb}{0.0, 0.5, 0.0}

\title{High-energy resummation in Higgs production at the next-to-leading order}
\author[a]{Francesco Giovanni Celiberto}
\author*[b]{Michael Fucilla}
\author[c]{Dmitry Yu. Ivanov}
\author[d,e]{Mohammed M.A. Mohammed}
\author[d,e]{Alessandro Papa}

\affiliation[a]{Universidad de Alcalá (UAH), Departamento de Física y Matemáticas, Campus Universitario, Alcalá de Henares, E-28805, Madrid, Spain}

\affiliation[b]{Université Paris-Saclay, CNRS/IN2P3, IJCLab, 91405, Orsay, France}

\affiliation[c]{Sobolev Institute of Mathematics, 630090 Novosibirsk, Russia}

\affiliation[d]{Dipartimento di Fisica, Università della Calabria, Arcavacata di Rende, I-87036, Cosenza, Italy}

\affiliation[e]{
INFN, Gruppo Collegato di Cosenza, Arcavacata di Rende, I-87036, Cosenza, Italy}

\emailAdd{francesco.celiberto@uah.es}
\emailAdd{michael.fucilla@ijclab.in2p3.fr}
\emailAdd{d-ivanov@math.nsc.ru}
\emailAdd{mohammed.maher@unical.it}
\emailAdd{alessandro.papa@fis.unical.it}

\abstract{We present the full next-to-leading order (NLO) result for the impact factor of a forward Higgs boson, obtained in the infinite-top-mass limit, both in the momentum representation and as superposition of the eigenfunctions of the leading-order (LO) BFKL kernel.}

\FullConference{16th International Symposium on Radiative Corrections: Applications of Quantum Field Theory to Phenomenology (RADCOR2023)\\
28th May - 2nd June, 2023\\
Crieff, Scotland, UK\\}

\begin{document}
\maketitle

\section{Introduction}

Precision physics in the Higgs sector has been one of the main challenges in recent years. The pure fixed-order calculations entering the \textit{collinear factorization} framework, which have been pushed up to the N3LO, are not able to describe the entire kinematic spectrum. In particular conditions, they must be necessarily supplemented by all-order \textit{resummations}; for instance, in the so called \textit{Regge} kinematic region, large energy-type logarithms spoil the perturbative behavior of the series and must be resummed to all orders. This resummation is, for instance, necessary to describe the inclusive hadroproduction of a forward Higgs in the limit of small Bjorken $x$, as well as to study inclusive forward emissions of a Higgs boson in association with a backward identified object. In Refs.~\cite{DelDuca:1993ga,DelDuca:2003ba}, pioneering studies were performed on the Higgs production in mini-jet events within the leading-logarithmic approximation (LLA). Phenomenological LLA analyses on the Higgs plus jet(s) production were performed within partial next-to-leading logarithmic approximation (NLLA) in Refs.~\cite{Celiberto:2020tmb,Andersen:2022zte}. High-energy effects from BFKL and Sudakov contributions were combined together to describe cross sections for the Higgs-plus-jet hadroproduction in almost back-to-back configurations~\cite{Xiao:2018esv}. Azimuthal correlations between a single-charmed hadron emitted in ultra-forward directions of rapidity and a Higgs boson were investigated within partial NLA in Ref.~\cite{Celiberto:2022zdg}.

Nevertheless, a complete resummation for these processes at full NLLA can be achieved through the Balitsky-Fadin-Kuraev-Lipatov (BFKL) approach~\cite{Fadin:1975cb,Kuraev:1976ge,Kuraev:1977fs,Balitsky:1978ic} (see Refs.~\cite{Boussarie:2017oae,Golec-Biernat:2018kem,Bolognino:2019yls,Celiberto:2020wpk,deLeon:2021ecb,Celiberto:2021dzy,Celiberto:2021fdp,Bolognino:2021niq,Celiberto:2022rfj,Celiberto:2022dyf,Celiberto:2022gji,Colferai:2023hje,Celiberto:2023rzw} for recent applications), but it requires the knowledge of the next-to-leading order Higgs impact factor. We present the full NLO result for the impact factor of a forward Higgs boson, obtained in the infinite-top-mass limit, both in the momentum representation and as superposition of the eigenfunctions of the LO BFKL kernel. 

\section{BFKL approach}
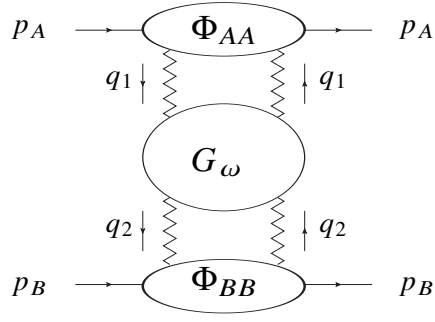
\begin{figure}
\begin{picture}(400,113)
\put(156,0){\includegraphics[width=0.27\textwidth]{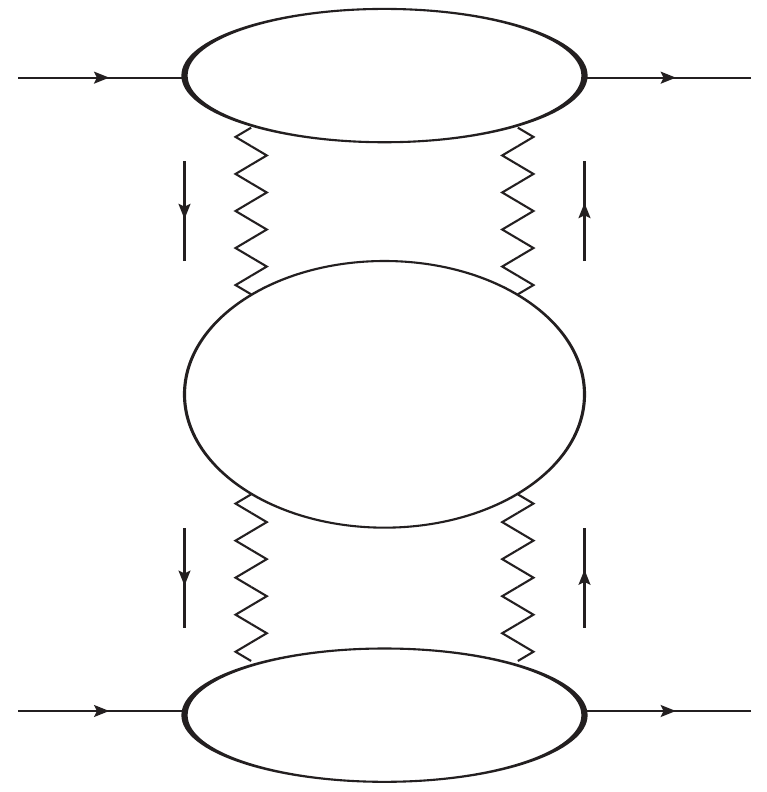}}
\put(202,105){\scalebox{1.2}{$\Phi_{AA}$}}
\put(202,55){\scalebox{1.2}{$G_{\omega}$}}
\put(202,10){\scalebox{1.2}{$\Phi_{BB}$}}
\put(135,10){$p_B$}
\put(135,107){$p_A$}
\put(280,10){$p_B$}
\put(280,107){$p_A$}
\put(170,32){$q_2$}
\put(170,88){$q_1$}
\put(250,32){$q_2$}
\put(250,88){$q_1$}
\end{picture}
\caption{Schematic representation of the factorized amplitude.}
\label{convolution}
\end{figure}
The BFKL equation is an integral equation that determines the behaviour at high energy $\sqrt{s}$ of the perturbative QCD amplitudes in which vacuum quantum numbers are
exchanged in the $t$-channel. It was derived in the LLA, which means collection of all terms of the type $\alpha_s^n \ln^n s$. This approximation leads to an increase of cross sections of the type
\begin{equation}
    \sigma_{\text{tot}}^{\text{LLA}} = \frac{s^{\omega_0}}{\sqrt{\ln s}} \; ,
\end{equation}
where $\omega_0 = \frac{g^2 C_A \ln 2}{\pi^2}$ is the rightmost singularity in the complex momentum plane of the $t$-channel partial wave
with vacuum quantum numbers (Pomerančuk singularity). \\

In the BFKL approach, using the $s$-channel unitarity relation, the imaginary part of a generic elastic scattering amplitude, $\mathcal{A}_{AB}^{AB}$, can be presented as\footnote{By making use of the optical theorem, this can be further related to total cross-sections.}
\begin{equation}
\Im_s \mathcal{A}_{AB}^{AB} = s \int_{\delta-i\infty}^{\delta+i\infty} \frac{d\omega}{2 \pi i} \int \frac{ d^{D-2} \vec{q}_{1} d^{D-2} \vec{q}_{2}}{{(2 \pi)^{D-2}}} \left( \frac{s}{s_0} \right)^{\omega} \frac{\Phi_A(\vec{q}_1)}{(\vec{q}_1^{\;2})^2}G_{\omega} (\vec{q}_1, \vec{q}_2) \frac{\Phi_B(-\vec{q}_2)}{(\vec{q}_2^{\;2})^2} ,
\label{Int:Eq:TotCross}
\end{equation}
where $G_{\omega}(\vec{q}_1,\vec{q}_2)$ is the Mellin transform of the BFKL Green's function, which satisfies the BFKL equation, $\Phi_A(\vec{q}_1)$ and $\Phi_B(\vec{q}_2)$ are the so-called \textit{impact factors}, $\delta$ is a real number which lies to the right of the right-most singularity of $G_{\omega}(\vec{q}_1,\vec{q}_2)$ and $s_0$ is scale introduced when performing the Mellin transform. \\

A schematic representation of the factorization formula~(\ref{Int:Eq:TotCross}) is given in Fig.~\ref{convolution}. Although the $s$-behavior is governed by the Green's function, the impact factors are necessary ingredients to construct the total amplitude. 
In this work, we will be interested in computing a NLO impact factor, for this reason, we report here the full next-to-leading definition:
\begin{gather}
\Phi_{AA}(\vec q_1; s_0) = \left( \frac{s_0}
{\vec q_1^{\:2}} \right)^{\omega( - \vec q_1^{\:2})}
\sum_{\{f\}}\int\theta(s_{\Lambda} -
s_{AR})\frac{ds_{AR}}{2\pi}\ d\rho_f \ \Gamma_{\{f\}A}^c
\left( \Gamma_{\{f\}A}^{c^{\prime}} \right)^* 
\langle cc^{\prime} | \hat{\cal P}_0 | 0 \rangle \nonumber \\ 
-\frac{1}{2}\int d^{D-2}q_2\ \frac{\vec q_1^{\:2}}{\vec q_2^{\:2}}
\: \Phi_{AA}^{(0)}(\vec q_2)
\: {\cal K}^{(0)}_r (\vec q_2, \vec q_1)\:\ln\left(\frac{s_{\Lambda}^2}
{s_0(\vec q_2 - \vec q_1)^2} \right)~ \; ,
\label{Int:Eq:ImpactproNext}
\end{gather}
where
\begin{equation}
    {\cal K}^{(0)}_r (\vec q_2, \vec q_1) = \frac{2 g^2 C_A}{(2\pi)^{D-1}} \frac{\vec{q}_1^{\; 2} \vec{q}_2^{\; 2} }{(\vec{q}_1-\vec{q}_2)^2}  
\label{Int:Eq:RealBornKert0}
\end{equation}
is the real part of the leading order BFKL kernel and
\begin{equation}
    \omega^{(1)}(t) = \frac{g^2 t}{(2 \pi)^{D-1}} \frac{N}{2} \int \frac{d^{D-2}k_{\perp}}{k_{\perp}^2 (q-k)_{\perp}^{2}} = - \frac{g^2 C_A \Gamma(1+\epsilon)(\vec{q}^{\; 2})^{-\epsilon}}{(4 \pi)^{2-\epsilon}} \frac{\Gamma^2(-\epsilon)}{\Gamma(-2\epsilon)} \; ,
    \label{ReggeTraj}
\end{equation}
is the one-loop Regge trajectory with $t=q^2 = - \vec{q}^{\; 2}$. The impact factor in Eq.~(\ref{Int:Eq:ImpactproNext}) is obtained by squaring off-shell amplitudes\footnote{The $t$-channel Reggeon is off-shell.}, $\Gamma_{\{f\}A}^c$, summing over all intermediate states, $\{ f \}$, and integrating over the phase-space of the intermediate particles, $d \rho_f$, and over the invariant mass $s_{AR}$ (invariant mass of the initial particle-Reggeon system). The factor $\langle cc^{\prime} | \hat{\cal P}_0 | 0 \rangle$ is necessary to project onto the color-singlet representation. It is important to mention that the first line in Eq.~(\ref{Int:Eq:ImpactproNext}) contains \textit{rapidity divergences} when the invariant mass $s_{AR}$ goes to infinity. From a technical point of view, the appearance of these divergences is due to the separation between the multi-Regge (MRK) and the quasi multi-Regge (QMRK) kinematics in the NLLA formulation of the BFKL approach. These divergences are regularized by the $\theta$-function in the first line of Eq.~(\ref{Int:Eq:ImpactproNext}) and then cancel with the second line in the same equation. This latter contribution comes exactly from the MRK contribution to the amplitude~(\ref{Int:Eq:TotCross}) in the NLLA. 
\section{Effective gluon–Higgs coupling and LO impact factor}
The calculation of the impact factor can be greatly simplified in the infinite-top-mass approximation. In this limit, we can employ the effective Lagrangian
\begin{equation}
\mathcal{L}_{ggH} = - \frac{g_H}{4} F_{\mu \nu}^{a} F^{\mu \nu,a} H \; , \hspace{1 cm} g_H = \frac{\alpha_s}{3 \pi v} \left( 1 + \frac{11}{4} \frac{\alpha_s}{\pi} \right) + {\cal O} (\alpha_s^3) \; ,
\label{EffLagrangia}
\end{equation}
which couples the Higgs field to gluons directly, via the QCD field strength tensor $F_{\mu \nu}^a = \partial_{\mu} A_{\nu}^a - \partial_{\nu} A_{\mu}^a  + g f^{abc} A_{\mu}^b A_{\nu}^c \;$.

At LO, the process is initiated by a collinear gluon that couples with the $t$-channel Reggeon to produce the forward Higgs boson. The gluon-initiated impact factor can be computed by using the definition given in Eq.~(\ref{Int:Eq:ImpactproNext}). Then, the proton-initiated impact factor is related to the gluon-initiated one by the factorization
 \begin{equation*}
    \frac{d \Phi_{PP}^{H}}{d x_H d^2 \vec{p}_H} = \int_{x_H}^1 \frac{d z_H}{z_H} f_g \left( \frac{x_H}{z_H} \right) \frac{d \Phi_{gg}^{H}}{d z_H d^2 \vec{p}_H} \; ,
 \end{equation*}
where $\vec{p}_H$ is the transverse momentum of the Higgs with respect to the collision axis, $z_H$ ($x_H$) is the longitudinal momentum fraction of the gluon (proton), and $f_g$ is the gluon parton distribution function (PDF). At LO, the forward-Higgs impact factor reads
\begin{equation}
    \frac{d \Phi_{PP}^{ \{ H \}(0)}}{d x_H d^2 \vec{p}_H} = \frac{g_H^2 \vec{q}^{\; 2} f_g (x_H) \delta^{(2)} ( \vec{q} - \vec{p}_H) }{8 \sqrt{N^2-1}} \; ,
\end{equation}
where $\vec{q}$ is the transverse momenta exchanged in the $t$-channel. This result correctly reproduces the large $m_t$-limit of the fully top-mass dependent result of Refs.~\cite{DelDuca:1993ga, Celiberto:2020tmb}.

In the next sections, we employ the effective Lagrangian in Eq.~(\ref{EffLagrangia}) to compute the NLO corrections to the Higgs impact factor in the infinite-top-mass limit.  

\section{Next-to-leading order result in the $k_T$-space}

\subsection{Real corrections}

At the next-to-leading order, the partonic sub-process can be initiated by a quark or a gluon. The quark-initiated contributions reads
\begin{gather}
      \frac{d \Phi_{q q}^{\{H q \}} (z_H, \vec{p}_H, \vec{q})}{d z_H d^2 \vec{p}_H} = \frac{\sqrt{C_A^2-1}}{16 C_A (2 \pi)^{D-1}} \frac{g^2 g_H^2}{(\vec{r}^{\; 2})^2} \left[ \frac{4 (1-z_H) \left( \vec{r} \cdot \vec{q} \; \right)^2 + z_H^2 \vec{q}^{\; 2} \vec{r}^{\; 2}}{z_H} \right] \; ,
\label{QuarkImp}
\end{gather}
while, the gluon-initiated one reads\footnote{This contribution does not contain the second line in Eq.~(\ref{Int:Eq:ImpactproNext}).}
\begin{gather}
\frac{d \Phi_{g g}^{\{H g \}} (\vec{q} \; )}{d z_H d^{2} \vec{p}_H} = \frac{g^2 g_H^2 C_A}{8 (2 \pi)^{D-1}(1-\epsilon) \sqrt{N^2-1}} \frac{2 \vec{q}^{\; 2}}{\vec{r}^{\; 2}} \nonumber \\
      \times \left[ \frac{z_H}{1-z_H} + z_H (1-z_H) + 2 (1-\epsilon) \frac{(1-z_H)}{z_H} \frac{(\vec{q} \cdot \vec{r})^2}{\vec{q}^{\; 2} \vec{r}^{\; 2}} \right]  \theta \left( s_{\Lambda} - s_{gR} \right) + \rm{finite \ terms} \; ,
     \label{GluonImp}
\end{gather}
where $\vec{r} = \vec{q} - \vec{p}_H$. In Eq.~(\ref{GluonImp}) we have shown only the phase-space singular part of the impact factor; the complete result can be found in Ref.~\cite{Celiberto:2022fgx}. There are three kinds of phase-space singularities in Eqs.~(\ref{QuarkImp}) and~(\ref{GluonImp}):
\begin{itemize}
    \item Rapidity divergences when $z_H \rightarrow 1$, present in the gluon-initiated contribution.
    \item Soft divergences when $\vec{r} \rightarrow 0$ and $z_H \rightarrow 1$, present in the gluon-initiated contribution.
    \item Collinear divergences when $\vec{r} \rightarrow 0$, present both in the gluon-initiated and in the quark-initiated contribution.
\end{itemize}
Results in Eqs.~(\ref{QuarkImp}) and~(\ref{GluonImp}) agree with ones in Ref.~\cite{Hentschinski:2020tbi}, independently performed in the Lipatov effective-action framework. 

\subsection{Virtual corrections}
Being impact factors obtained by squaring effective vertices, we must extract the $1$-loop effective vertex for the production of a Higgs in gluon-Reggeon collisions. To this aim, we use a reference amplitude and compare it with the expected Regge form. We employ the amplitude for the diffusion of a gluon off a quark to produce a Higgs plus a quark, $\mathcal{A}_{gq \rightarrow Hq}^{(8,-)}$, with octet color state and negative signature in the $t$-channel. It should assume the following Reggeized form\footnote{The apexes $(0)$ and $(1)$ denote the Born and the $1$-loop approximation, respectively.}
$$
{\cal A}_{g q \rightarrow H q}^{(8,-)} = \Gamma_{ H g}^{ac} \frac{s}{t}\left[ \left( \frac{s}{-t} \right)^{\omega(t)} + \left(
\frac{-s}{-t} \right)^{\omega(t)} \right]\Gamma_{qq}^{c} \approx
\Gamma_{H g}^{ac(0)} \frac{2s}{t} \Gamma_{qq}^{c(0)}
$$
\begin{equation}
+ \Gamma_{H g}^{ac(0)} \frac{s}{t}\omega^{(1)}(t)
\left[ \ln\left( \frac{s}{-t} \right) + \ln\left(
\frac{-s}{-t} \right) \right]\Gamma_{q q}^{c(0)} + \Gamma_{H g}^{ac(0)} \frac{2s}{t}\Gamma_{q q}^{c(1)} +
\Gamma_{H g}^{ac(1)}\frac{2s}{t} \Gamma_{qq}^{c(0)} \; ,
\label{ReggeFormEx1}
\end{equation}
where $\omega (t)$ is the Regge trajectory, $ \Gamma_{ H g}^{ac}$ is the LO gluon-Higgs-Reggeon effective vertex and lastly $\Gamma_{qq}^{c}$ is the quark-quark-Reggeon effective vertex.  Since the only unknown ingredients in the right-hand side of Eq.~(\ref{ReggeFormEx1}) is the one-loop correction to the gluon-Higgs-Reggeon vertex, if we compute the amplitude ${\cal A}_{g q \rightarrow H q}^{(8,-)}$, we are immediately able to extract it. The effective vertex allows us to obtain the virtual contribution to the impact factor, which reads\footnote{For further details about the computation, see~\cite{Celiberto:2022fgx,Fucilla:2022whr,Celiberto:2023dkr_article}.}
\begin{equation*}
\frac{d \Phi_{gg}^{\{ H \}(1)}}{d z_H d^2 \vec{p}_H} = \frac{d \Phi_{gg}^{\{ H \}(0)}}{d z_H d^2 \vec{p}_H} \; \frac{\bar{\alpha}_s}{2 \pi} \left( \frac{\vec{q}^{\; 2}}{\mu^2}  \right)^{- \epsilon}  \left[  - \frac{C_A}{\epsilon^2} + \frac{11 C_A - 2n_f}{6 \epsilon} \right.
\end{equation*}
\begin{equation}
      \left. - \frac{C_A}{\epsilon} \ln \left( \frac{\vec{q}^{\; 2}}{s_0} \right) - \frac{5 n_f}{9} + C_A \left( 2\;\Re \left( {\rm{Li}}_2 \left( 1 + \frac{m_H^2}{\vec{q}^{\; 2}} \right)\right) + \frac{\pi^2}{3} + \frac{67}{18} \right) + 11 \right] \; .
      \label{VirtualPartIMF}
\end{equation}
In the computation $\epsilon=\epsilon_{UV}=\epsilon_{IR}$ is set such that any scaleless Feynman integral does not contribute to virtual corrections. The result in Eq.~(\ref{VirtualPartIMF}) is compatible with the independent result of Ref.~\cite{Nefedov:2019mrg}, performed within the Lipatov effective action framework\footnote{To compare the two results, a reference physical amplitude must be considered; it must be reconstructed in both approaches. This is because there is freedom in how individual impact factors are defined.}.

\section{Next-to-leading order result in the ($n, \nu$)-space and cancellation of divergences}

\subsection{The ($n, \nu$)-space result}
The cancellation of divergences can be seen only performing the integration of transverse momenta in Eq.~(\ref{Int:Eq:ImpactproNext}). Nevertheless, in order to avoid a complete convolution between two impact factors and the Green function in the $k_T$-space, which would be complicated and would not lead to a fully general result\footnote{It would depend on the target impact factor.}, one can move to the ($n, \nu$)-space. We explain the procedure, in the LLA, for pedagogical purpose. The BFKL Green function in Eq.~(\ref{Int:Eq:ImpactproNext}) can be represent trough a spectral representation onto the eigenfunctions of its LO kernel, as
\begin{equation}
    G_{\omega }^{\left( 0\right) }\left( \vec{q}_{1},\vec{q}_{2}\right) = \sum_{n=-\infty}^\infty \int_{-\infty}^{+\infty} d \nu \frac{\phi_{\nu}^{n} (\vec{q}_1^{\; 2}) \phi_{\nu}^{n *}(\vec{q}_2^{\; 2})}{\omega - \frac{\alpha_s C_A}{\pi} \chi(n, \nu)} \; ,
\end{equation}
where $\phi_{\nu}^{n} (\vec{q}^{\; 2})$ are the LO BFKL kernel eigenfunctions and $(\alpha_s C_A/\pi) \chi(n, \nu)$ the corresponding eigenvalues, with
\begin{equation}
    \phi_{\nu}^{n} (\vec{q}^{\; 2}) = \frac{1}{\pi \sqrt{2}} (\vec{q}^{\; 2})^{i \nu - \frac{1}{2}} e^{i n \phi} \; , \hspace{0.5 cm} \chi(n, \nu) = 2 \psi (1) - \psi \left( \frac{n}{2} + \frac{1}{2} + i \nu \right) - \psi \left( \frac{n}{2} + \frac{1}{2} - i \nu \right) \; .
\end{equation}
Now, each impact factor integrate separately with one eigenfunction and we can give the definition of the ($n, \nu$)-space projected impact factor,   
\begin{equation}
     d \Phi^{(0)}_{A A}(n,\nu) \equiv \int \frac{d^{2-2\epsilon} q}{\pi \sqrt{2}} (\vec{q}^{\; 2})^{i \nu - \frac{3}{2}} e^{i n \phi} d \Phi _{A A}^{\left(0\right)} (\vec{q} \;)  \; .
\label{ProjectionDef}
\end{equation}
Then, the projected LO Higgs impact factor reads  
\begin{equation}
    \frac{d \Phi_{ PP }^{\{H \}(0)} (x_H, \vec{p}_H, n, \nu)}{d x_H d^2 \vec{p}_H}  = \frac{g_H^2}{8 (1-\epsilon) \sqrt{N^2-1}} \frac{(\vec{p}_H^{\; 2})^{i \nu - \frac{1}{2}} e^{i n \phi_H}}{\pi \sqrt{2}} f_g (x_H)  \; .
\end{equation}
At NLO, the integration in (\ref{ProjectionDef}) convert phase-space singularities of reals corrections in $\epsilon$-poles, allowing to observe the cancellation of divergences. 
\subsection{Cancellation of divergences}
Individual partonic contributions to the impact factor are clearly divergent. In order to build a finite quantity, we have to show the explicit cancellation of these divergences. We start discussing the rapidity divergences typical of high-energy computations. These latter are generated by the separation between MRK and QMRK; they are only present in the gluon-initiated contribution and give a $\ln s_{\Lambda}$-term after integration over the phase space. For this reason, the $d \Phi_{P P}^{\{H g \}}$ contribution must be combined with the second line in Eq.~(\ref{Int:Eq:ImpactproNext}). Symbolically, we then have 
\begin{equation}
      d \tilde{\Phi}_{P P}^{\{H g \}} = d \Phi_{P P}^{\{H g \}} - d \Phi_{P P}^{\{H \}} \otimes {\mathcal{K}_r^{(0)} \ln s_{\Lambda}} \; ,
\end{equation}
where $d \tilde{\Phi}_{P P}^{\{H g \}}$ is free from rapidity divergences. This cancellation leave us with a double $\epsilon$-pole singularity, coming from the kinematical region where $z_H \sim 1$ and $\vec{r} \sim \vec{0}$, after the integration over phase-space.

Virtual corrections are affected by UV-divergences, which can be removed performing the renormalization of the strong coupling, {\it i.e.} 
\begin{equation}
    \alpha_s (\mu^2) = \alpha_s (\mu_R^2) \left[ 1 + \frac{\alpha_s (\mu_R^2)}{2 \pi} \beta_0 \left( -\frac{1}{\epsilon} - \ln (4 \pi e^{-\gamma_E}) + \ln \left( \frac{\mu_R^2}{\mu^2} \right) \right) \right] \; .
    \label{alpha_Run}
\end{equation}
Soft divergences should cancel in the real plus virtual combination, but, some IR-singularity (of collinear nature), survive this cancellation. They are initial state divergences that, within this scheme\footnote{We stress that we are treating the IR and UV divergences by adopting the same regulator.}, should be cancelled by renormalizing the gluon PDF, {\it i.e.} 
\begin{gather}
    f_g (x, \mu) = f_g (x, \mu_F) - \frac{\alpha_s(\mu_F)}{2 \pi} \left( - \frac{1}{\epsilon} - \ln (4 \pi e^{- \gamma_E}) + \ln \left( \frac{\mu_F^2}{\mu^2} \right) \right) \nonumber \\
    \times \int_x^1 \frac{dz}{z} \left[ P_{gq} (z) \sum_{a=q \bar{q}} f_a \left( \frac{x}{z} , \mu_F \right) + P_{gg} (z) f_g \left( \frac{x}{z} , \mu_F \right) \right] \; .
     \label{fg_Run}
\end{gather} 
The cancellation of divergences takes place and the complete finite result can be expressed in terms of integrals of hypergeometric functions. We refer to the original work~\cite{Celiberto:2022fgx} for the complete result.

\section{Conclusions and outlook}
We calculated the full NLO correction to the impact factor for the production of a Higgs boson emitted by a proton in the forward rapidity region. 
Its analytic expression was obtained both in the momentum and in the Mellin representations. The latter is particularly relevant to clearly observe a complete cancellation of NLO singularities, and it is useful for future numerical studies~\cite{Celiberto:2023uuk_article}.

\begingroup
\setstretch{1.0}
\bibliographystyle{apsrev}
\bibliography{references}
\endgroup
\end{document}